\documentclass[aps,twocolumn,showpacs]{revtex4}
\usepackage{epsfig}
\usepackage{amsbsy}
\usepackage{amsmath}
\usepackage{amsfonts}
\usepackage{graphicx}
\usepackage{latexsym}
\def\CC{{\rm\kern.24em \vrule width.04em height1.46ex depth-.07ex
\kern-.30em C}}
\begin{document}
\title{Relations between bosonic quadrature squeezing and atomic spin squeezing}
\author{Xiaoguang Wang$^{1,3}$ and Barry C Sanders$^2$}
\affiliation{1. Department of Physics and Australian Centre of Excellence for Quantum
Computer Technology, \\
Macquarie University, Sydney, New South Wales 2109, Australia.}
\affiliation{2. Quantum Information Science Group, Department of Physics and Astronomy, University of Calgary, Alberta T2N 1N4, Canada.}
\affiliation{3. Department of Physics and Center for Nonlinear Studies, Hong Kong Baptist University, Hong Kong, China.}

\date{\today}
\begin{abstract}
We study relations between bosonic quadrature squeezing and atomic spin squeezing, and find that the latter reduces to the former in the limit of a large number of atoms for even and odd states. 
We demonstrate this reduction by treating even and odd spin coherent states, for which analytical solutions are readily obtained, and prove that even spin coherent states always exhibit spin squeezing, whereas odd spin coherent states do not, analogous to the squeezing characteristic of even and odd bosonic coherent states. 
Finally, we examine the squeezing transfer between photons and atoms via the Dicke Hamiltonian, where a perfect transfer of squeezing is demonstrated in the limit of a large number of atoms.
\end{abstract}
\maketitle

\section{introduction}
The investigation of squeezed states of light~\cite{Squ} is of fundamental interest, and has applications to practical precision measurements~\cite{Cave81,Bond84}. 
Recently, spin squeezed states for an ensemble of atoms~\cite
{Wod85,Kitagawa,Wineland,Agarwal,Lukin,Vernac,Kuzmichqnd,Sorensen1,Youli,hald,kasevich,uffe,Dominic,Usha,Gasenzer,Stockton,Law,Pu1,WangSpin,WangSuper,Rojo} have become important, with potential applications to atomic interferometers and  precise atomic clocks. 
For quantum information applications~\cite{Nie00}, the close relation between
atomic spin squeezing and quantum entanglement~\cite{Sorensen01,Dominic2,Sanmore,Kitagawa2,WangNew} enhance the 
importance of atomic spin squeezing.
Entanglement and bosonic squeezing are also related~\cite{Ephoton}. Here, we investigate the relationship between these two mathematically distinct yet intuitively connected squeezing figures of merit.

Let us first review bosonic quadrature squeezing by introducing two quadrature operators $X$ and $P$ given by
\begin{equation}
X=a+a^{\dagger }, \quad P=-i(a-a^{\dagger }), 
\end{equation}
where operators $a$ and $a^\dagger$ are the annihilation and creation operators of a boson, respectively. The two quadrature operators satisfy the commutation relation $[X,P]=2i$, which yields the uncertainty relation $(\Delta X)^2(\Delta P)^2\ge 1$. A state is squeezed if either $(\Delta X)^2$ or $(\Delta P)^2$ is less than 1. 

Second, we review the atomic spin squeezing by considering an ensemble of $N$ two-level atoms characterized by an angular
momentum vector operator $\mathbf{S}=(S_x,S_y,S_z)$ satisfying the su(2)
algebra
\begin{equation}
\lbrack S_z,S_{\pm }]=\pm S_{\pm },\,[S_{+},S_{-}]=2S_z,\,S_{\pm }=S_x\pm iS_y.
\label{su2}
\end{equation}
Several definitions of spin squeezed states permeate in the 
literature ~\cite{Wod85,Kitagawa,Wineland,Sorensen01,Usha03}. 
There are two well-accepted definitions, one is 
given by Kitagawa and Ueda~\cite{Kitagawa}, and another by Wineland et al.~\cite{Wineland}. The spin squeezing parameter associated with the former definition is given by~\cite{Kitagawa}
\begin{equation}
\xi =\frac{2(\Delta S_{\mathbf{n}_{\bot }})^2}j=\frac{2\langle S_{\mathbf{n}_{\bot
}}^2\rangle }j. \label{xixixi}
\end{equation}
where the subscript ${\bf n}_\perp $ refers to an axis perpendicular to the mean spin $\langle{\bf S}\rangle$, for which the minimal value of the variance $(\Delta S)^2$ is obtained, $j=N/2$, and $S_{{\bf n}_\perp}={\bf S}\cdot {\bf n}_{\perp}$. The inequality $\xi<1$ indicates that the system is spin squeezed. We will also consider the definition of spin squeezing given by Wineland et al.~\cite{Wineland}, and show that the spin squeezing by the two definitions reduces to the bosonic squeezing in the limit of large number of atoms. Another reason for adopting the definitions is that they are both closely related to quantum entanglement of $N$ two-level atoms~\cite{Sorensen01,Dominic2,Sanmore,Kitagawa2,WangNew}.

It is well-known that the Heisenberg-Weyl algebra describing the bosonic mode can be obtained by contraction from the su(2) algebra describing the ensemble of atoms~\cite{Chumakov}. To see this, we define $%
b\equiv S_{-}/\sqrt{2j}$ and $b^{\dagger }\equiv S_{+}/\sqrt{2j}.$ From the
commutation relation (\ref{su2}), we have
\begin{equation}
\lbrack {\cal N},b^{\dagger }]=b^{\dagger },\;[{\cal N}%
,b]=-b,\;[b,b^{\dagger }]=1-\frac{{\cal N}}j, 
\end{equation}
where ${\cal N}=S_z+j$ is the ``number operator", and its eigenvalues vary from 0 to $N$ counting the number of excited atoms.  
In the limit of $j\rightarrow \infty ,$ the operators ${\cal N},$ $b,$
and $b^{\dagger }$ satisfy the commutation relations of the Heisenberg-Weyl
algebra. Note that, when we take the limit, the average number of excited atoms $\langle{\cal N}\rangle$ should be much less than the total number of atoms $N$.

We can also use the usual Holstein-Primakoff transformation~\cite{Emary}:
\begin{equation}
S_{+}=a^{\dagger }\sqrt{2j-a^{\dagger }a},\,S_{-}=\sqrt{2j-a^{\dagger }a}%
a,\,S_z=a^{\dagger }a-j. 
\end{equation}
In the limit of $j\rightarrow \infty ,$ we have
\begin{equation}
\frac{S_{+}}{\sqrt{2j}}\rightarrow a^\dagger,\;
\frac{S_{-}}{\sqrt{2j}}\rightarrow a,\;-\frac{S_z}j\rightarrow 1, 
\label{limit2}
\end{equation}
by expanding the square root and neglecting terms of $O(1/j)$. 
We see that the bosonic system and the atomic spin system are connected by the large $j$ limit from an algebraic point of view. 
This connection motivates us to ask if these two kinds of squeezing are also related by this limit. 
Moreover, the link between bosonic squeezing and atomic squeezing is exemplified by the fact that the squeezed light can be obtained from spin squeezed atoms by Raman scattering of a strong laser pulse~\cite{uffe}, implying a close relation between these two kinds of squeezing. 
We also study squeezing transfer between the light field and the atomic system. These analyses points to a relation between bosonic squeezing and atomic squeezing that helps to resolve the ambiguity of defining spin squeezing.

The paper is organized as follows. In Sec.~II, we show that atomic spin squeezing reduces to bosonic quadrature squeezing in the large $N$ limit. As an example, in Sec.~III, even and odd spin coherent states are considered, and spin squeezing of them exactly reduces to the squeezing of even and odd bosonic coherent states in the above limit. In Sec.~IV, we study squeezing transfer between light field and an atomic system. We conclude in Sec.~V.

\section{Relation between bosonic squeezing and atomic squeezing}
In the standard analyses of bosonic quadrature squeezing, both the variances $(\Delta X)^2$ and $(\Delta P)^2$ are calculated, 
and $(\Delta X)^2<1$ or $(\Delta P)^2<1$ indicates bosonic squeezing. 
Here, for our purposes, we consider the so-called principle quadrature squeezing~\cite{Psqu}, and define an appropriate quadrature operator
\begin{equation}
X_\theta=X\cos\theta+P\sin\theta=a e^{-i\theta}+a^\dagger e^{i\theta}
\end{equation}
with $X=X_0$ and $P=X_{\pi/2}$ being special cases. Bosonic squeezing is characterized by one parameter~\cite{Psqu} 
\begin{equation}
\zeta =\min_{\theta\in [0,2\pi) }(\Delta X_\theta)^2,
\label{squ1}
\end{equation}
which is the minimum value of $(\Delta X_\theta)^2$ with respect to $\theta$, and $\zeta<1$ indicates principle squeezing. 
The definition of $\zeta$ provides an atomic squeezing counterpart to bosonic squeezing.

To display this connection, we consider even (odd) states. These states refer to those being a superposition of even (odd) Fock states for the bosonic system, and those being a superposition of Dicke states $|n\rangle _j\equiv |j,-j+n\rangle$ with the even (odd) excitations for the atomic system.    
The Dicke states $|n\rangle_j $ satisfies~\cite
{Dicke}
\begin{equation}
\quad {\cal N}|n\rangle _j=n|n\rangle _j.
\end{equation}
Specifically, even and odd bosonic coherent states have been realized experimentally for a quantized cavity field~\cite{Brune96}, vibrational motion of trapped ions~\cite{Monroe96}, and an electron in a Rydberg atomic system~\cite{Noel96}. In the context of cavity QED, Gerry and Grobe~\cite{Gerry97} showed that the even and odd atomic spin coherent states can be generated by the use of a dispersive interaction between atoms and field via a state reduction technique. The basis idea of the technique is that we first entangle atoms and field, and then make an appropriate quantum measurement. After the measurement, we can obtain desire states such as even and odd states.

Even and odd states are considered here because, in addition to exhibiting 
squeezing, such states are amenable to exact analytical calculations. Thus, these states serve as examples for demonstrating connections between bosonic and atomic squeezing.

For even (odd) states, $\langle a\rangle=0$, so, from Eq.~(\ref{squ1}), we obtain
\begin{align}
\zeta =&\min_\theta \langle X_\theta ^2\rangle 
\nonumber \\
=&\min_\theta \left[ \cos ^2\theta \langle X^2\rangle +\sin ^2\theta
\langle P^2\rangle +\frac{\sin (2\theta )}2\langle XP+PX\rangle \right] 
\nonumber \\
=&1+2\langle a^\dagger a\rangle -2|\langle a^2\rangle | 
=\langle a^{\dagger }a+aa^{\dagger }\rangle -2|\langle a^2\rangle |.
\label{squu1}
\end{align}
The above equation implies a sufficient condition for bosonic squeezing 
\begin{equation}
\tilde{\zeta}\equiv\langle a^\dagger a\rangle -|\langle a^2\rangle |<0. 
\end{equation}
Obviously, one necessary condition for squeezing is that $|\langle a^2\rangle|\neq 0$.

For even (odd) atomic states, the squeezing parameter $\xi$ is given by~\cite{WangNew}
\begin{eqnarray}
\xi &=&1+j-\frac 1j\left[ \langle S_z^2\rangle +|S_{-}^2|\right]  \nonumber
\\
&=&1+2\langle {\cal N}\rangle -\frac{\langle {\cal N}^2\rangle }j-%
\frac{|\langle S_{-}^2\rangle |}j  \nonumber \\
&=&\frac{\langle S_{+}S_{-}+S_{-}S_{+}\rangle }{2j}-\frac{|\langle
S_{-}^2\rangle |}j  \label{squ2},
\end{eqnarray}
from which we obtain a sufficient condition for atomic squeezing 
\begin{equation}
\tilde{\xi}\equiv 2j\langle {\cal N}\rangle -\langle {\cal N}^2\rangle
-|\langle S_{-}^2\rangle |<0.  \label{criteria2}
\end{equation}

Using Eq.~(\ref{limit2}), in the limit of $j\rightarrow\infty$, we find that 
Eq.~(\ref{squ2}) reduces to Eq.~(\ref{squu1}) for even (odd) states. 
This result displays a direct connection between bosonic squeezing and atomic squeezing. From an experimental point of view, the number of atoms is typically large enough, so the observed atomic squeezing is expected to approximate the bosonic  quadrature squeezing. As a remark, 
Eqs.~(\ref{squu1}) and (\ref{squ2}) obtained for even and odd states are also applicable to arbitrary states, which is discussed in Appendix A.

\section{Even and odd coherent states}
We now exemplify the reduction displayed above by first considering the even bosonic coherent state (EBCS) $|\alpha \rangle _{+}$ and the odd bosonic coherent state (OBCS) $|\alpha \rangle _{-}$~\cite{EOCS}, and then even spin coherent state (ESCS) and odd spin coherent state (OSCS) of the atomic system~\cite{Gerry97,WangSuper}.

The EBCS and OBCS are defined as~\cite{EOCS} 
\begin{align}
|\alpha \rangle _{\pm }=&\frac 1{\sqrt{2[1\pm \exp (-2|\alpha |^2)]}}\left(
|\alpha \rangle \pm |-\alpha \rangle \right)\nonumber\\
=&\frac {e^{-|\alpha|^2/2}}{\sqrt{2[1\pm \exp (-2|\alpha |^2)]}}\sum_{n=0}^{\infty}\frac{\alpha^n[1\pm(-1)^n]}{\sqrt{n!}},
\label{eocs}
\end{align}
where 
\begin{equation}
|\alpha\rangle=e^{-|\alpha|^2/2}\sum_{n=0}^{\infty}\frac{\alpha^n}{\sqrt{n!}}|n\rangle,\quad \alpha \in \CC 
\end{equation}
is the usual bosonic
coherent state (BCS). The BCS, EBCS, and OBCS satisfy
\begin{equation}
a|\alpha \rangle =\alpha |\alpha \rangle,\quad a^2|\alpha \rangle _{\pm }=\alpha
^2|\alpha \rangle _{\pm },
\end{equation}
from which we see that $_{\pm }\langle \alpha |a^2|\alpha \rangle _{\pm
}=\alpha ^2.$ From Eq.~(\ref{squu1}), in order to calculate the squeezing
parameter $\zeta$, we need to know the mean value of $a^\dagger a$, which is given by
\begin{equation}
_{\pm }\langle \alpha |a^\dagger a|\alpha \rangle _{\pm }=\frac{1\mp \exp (-2|\alpha
|^2)}{1\pm \exp (-2|\alpha |^2)}|\alpha |^2.
\end{equation}
Therefore, the squeezing parameters for the EBCS and OBCS are 
\begin{eqnarray}
\zeta _{+} &=&1+2|\alpha |^2[\tanh (|\alpha |^2)-1]<1, \\
\zeta _{-} &=&1+2|\alpha |^2[\coth (|\alpha |^2)-1]>1, \label{zetapm}
\end{eqnarray}
repectively. Then, we recover the result that the EBCS is always squeezed,
whereas the OBCS is not squeezed~\cite{EOCS}

We introduce the ESCS and OSCS as 
\begin{align}
|\eta \rangle _{\pm } =&\frac 1{\sqrt{2\pm 2\gamma ^{2j}}}\left( |\eta
\rangle \pm |-\eta \rangle \right)   \nonumber \\
=&\frac{(1+|\eta |^2)^{-j}}{\sqrt{2\pm 2\gamma ^{2j}}}\sum_{n=0}^{2j}{{{%
\binom{2j}n}}}^{1/2}\eta ^n
[1\pm (-1)^n]|n\rangle _j,  \label{eq:state}
\end{align}
where $\gamma =\frac{1-|\eta |^2}{1+|\eta |^2}\in (0,1)$, and 
$|\eta\rangle$ denotes the spin coherent state (SCS)~\cite{SCS}, 
\begin{equation}
|\eta \rangle =(1+|\eta |^2)^{-j}\sum_{n=0}^{2j}{{{\binom{2j}n}}}^{1/2}\eta
^n|n\rangle _j, \; \eta\in\CC.
\end{equation}
with the parameter $\eta $ being complex. Due to the fact that the probability
distribution $|_j\langle n|\eta \rangle |^2$ is a binomial distribution, we
restrict $|\eta |\in (0,1).$ 

Next, we examine squeezing properties of the ESCS and OSCS, for which we have\\
{\bf Proposition 1}: \emph{The even spin coherent state is always spin squeezed,
whereas the odd spin coherent state is never squeezed.}\\
{\it Proof}: First we show that the spin squeezing parameter $\xi $ can be
simplified for the ESCS and OSCS. 
From the definition of the SCS, it is
straightforward to check that 
\begin{equation}
S_{-}|\eta \rangle =\eta (2j-{\cal N})|\eta \rangle,  \label{ladder}
\end{equation}
from which we obtain
\begin{equation}
S_{-}^2|\eta \rangle _{\pm }=\eta ^2(2j-{\cal N})(2j-{\cal N}-1)|\eta
\rangle _{\pm }.  \label{ladder2}
\end{equation}
Substituting the above equation into Eq.~(\ref{squ2}) leads to the result that the spin squeezing parameter is only determined by the expectation values $\langle 
{\cal N}\rangle $ and $\langle {\cal N}^2\rangle .$ Another result is that $\langle
S_{-}^2\rangle /\eta ^2\geq 0.$ To obtain this result, we first notice
that if the ESCS and OSCS are written as 
\begin{equation}
|\eta \rangle _{\pm
}=\sum_{n=0}^{2j}\eta ^nC_{n,\pm }|n\rangle _j, 
\end{equation}
then $C_{n\pm }\geq 0$, and  
\begin{align}
\langle S_{-}^2\rangle =&\eta ^2\sum_{n=0}^{2j-2}|\eta |^{2n}C_{n,\pm
}C_{n+2,\pm }\nonumber\\
&\times\sqrt{(n+2)(n+1)(2j-n)(2j-n-1)}.
\end{align}
Since $C_{n\pm }\geq 0,$ from the above equation, we obtain $\langle
S_{-}^2\rangle /\eta ^2\geq 0.$ Thus, from Eq.~(\ref{ladder2}), we find 
\begin{equation}
\langle (2j-{\cal N})(2j-{\cal N}-1)\rangle \geq 0.  \label{fact}
\end{equation}

By using Eq.~(\ref{fact}), the spin squeezing parameter $\xi $ of Eq.~(\ref{squ2})
simplifies to
\begin{align}
\tilde{\xi} =&[2j+(4j-1)|\eta |^2]\langle {\cal N}\rangle -(1+|\eta
|^2)\langle {\cal N}^2\rangle \nonumber\\
&-2j(2j-1)|\eta |^2  \nonumber \\
=&[2j-1+(4j-2)|\eta |^2]F_1-(1+|\eta |^2)F_2\nonumber\\
&-2j(2j-1)|\eta |^2,
\label{xibar}
\end{align}
where $F_1=\langle {\cal N}\rangle $ and $F_2=\langle \mathcal{N(N}-1%
\mathcal{)}\rangle $ are the factorial moments which are introduced for
convenience of the following discussions. So, we see that the spin squeezing
parameter is expressed as a linear combination of $\langle {\cal N}%
\rangle $ and $\langle {\cal N}^2\rangle ,$ or equivalently, of $F_1$ and $%
F_2.$

Let us consider the ESCS. The associated factorial moments are given by~\cite
{WangSuper} 
\begin{align}
F_1=&\frac{2j|\eta |^2}{1+|\eta |^2}\left( \frac{1-\gamma ^{2j-1}}{1+\gamma
^{2j}}\right),\nonumber\\
F_2=&\frac{2j(2j-1)|\eta |^4}{(1+|\eta |^2)^2}\left( \frac{%
1+\gamma ^{2j-2}}{1+\gamma ^{2j}}\right) ,\label{fff}
\end{align}
which obey the inequalities 
\begin{equation}
F_1<\frac{2j|\eta |^2}{1+|\eta |^2},\quad
F_2>\frac{2j(2j-1)|\eta |^4}{(1+|\eta
|^2)^2}.  \label{ine1}
\end{equation}
The above equation results directly from the fact that $\gamma \in (0,1).$
Applying Eq.~(\ref{ine1}) to Eq.~(\ref{xibar}), we obtain
$
\tilde{\xi}<0.
$
Therefore, the ESCS is always spin squeezed. 

For the OSCS, the associated factorial moments are given by
~\cite{WangSuper} 

\begin{align}
F_1=&\frac{2j|\eta |^2}{1+|\eta |^2}\left( \frac{1+\gamma ^{2j-1}}{1-\gamma
^{2j}}\right),\nonumber\\
F_2=&\frac{2j(2j-1)|\eta |^4}{(1+|\eta |^2)^2}\left( \frac{%
1-\gamma ^{2j-2}}{1-\gamma ^{2j}}\right) ,
\end{align}
which obey

\begin{equation}
F_1>\frac{2j|\eta |^2}{1+|\eta |^2},\quad
F_2<\frac{2j(2j-1)|\eta |^4}{(1+|\eta
|^2)^2}.  \label{ine2}
\end{equation}
Applying Eq.~(\ref{ine2}) to Eq.~(\ref{xibar}) leads to  $\tilde{\xi}>0$
indicating that the OSCS is never spin squeezed.{$\square $}

Now we display the reduction of the atomic spin squeezing to bosonic squeezing for ESCS and OSCS. From Eqs.~(\ref{squ2}) and (\ref{criteria2}), we read that 
\begin{equation}
\xi=1+\frac{\tilde{\xi}}{j}. 
\end{equation}
For ESCS and OSCS, the quantity $\tilde{\xi}$ is given by 
Eq.~(\ref{xibar}). Now we take the limit $j\rightarrow \infty, \,|\eta|\rightarrow 0$, keeping $2j|\eta|^2=|\alpha|^2$ fixed. In this limit, from Eq.~(\ref{fff}), we see that $F_1\rightarrow |\alpha|^2\tanh(|\alpha|^2), \,F_2\rightarrow |\alpha|^4$. Therefore, we obtain
\begin{equation}
\frac{\tilde{\xi}}{j}\rightarrow 2|\alpha|^2[\tanh(|\alpha|^2)-1],
\end{equation} 
and $\xi$ reduces to $\zeta_+$ (\ref{zetapm}) in this limit. Similarly, we can show that the squeezing parameter $\xi$ reduces to $\zeta_-$ for OSCS. 
Recognizing the close relation between bosonic squeezing and atomic squeezing, we next study the squeezing transfer between light and an atomic system.

\section{Squeezing transfer}
Atomic squeezed states can be generated by interacting with squeezed light~\cite{Agarwal,Kuzmich}. Conversely, squeezed light states can be generated from squeezed atoms~\cite{Saito}. Here, we study the squeezing transfer from 
light to atomic system via the single-mode Dicke Hamiltonian under the rotating wave approximation ($\hbar=1$)~\cite{Dicke,Tavis} 
\begin{equation}
H=\omega_0 S_z +\omega a^\dagger a +\frac{\lambda}{\sqrt{2j}}(S_+a+S_-a^\dagger),
\end{equation}
where $\omega_0$ is the atomic splitting, $\omega$ is the field frequency, and $\lambda$ is the atom-field coupling. For the resonant case ($\omega=\omega_0$),
the Dicke Hamiltonian in the interaction picture is written as
\begin{equation}
H_{\rm I}=\frac{\lambda}{\sqrt{2j}}(S_+a+S_-a^\dagger).
\end{equation}
In the following, we work in the interaction picture. 

There is a conserved parity $\Pi$ associated with the Dicke Hamiltonian, which is given by
\begin{equation}
\Pi=(-1)^{{\cal N}+a^\dagger a}, \quad [\Pi,H]=[\Pi,H_{\rm I}]=0.
\end{equation}
We choose the initial state of the whole system as
\begin{equation}
|\psi(0)\rangle=|0\rangle_j\otimes |\alpha_0\rangle_{+};
\end{equation}
i.e., the light is in the EBCS~(\ref{eocs}) and each atom is prepared in the ground state. Here, in order to see better squeezing transfer, we choose $\alpha_0=0.7995$ which corresponds to maximal squeezing ($\zeta=0.4431$) of the EBCS with respect to $\alpha$. The state vector 
at time $t$ under Hamiltonian evolution is formally written as
\begin{equation}
|\psi(t)\rangle=e^{-iH_{\rm I}t}|\psi(0)\rangle.
\end{equation}
It is easy to see that the state at arbitrary time $t$ has a definite parity 1, i.e., $\Pi |\psi(t)\rangle=|\psi(t)\rangle$. This result leads to the expectation values of $S_\pm$, $a$, and $a^\dagger$ being zero since the parity will change after the action of these operators on the state. Thus, for state $|\psi(t)\rangle$, we may apply Eqs.~(\ref{squu1}) and (\ref{squ2}) to 
examine  squeezing.

Before numerical calculations of squeezing, we consider two special cases: $N=1$ and the limit $N\rightarrow\infty$. For $N=1$, the single two-level atom cannot be spin squeezed since $\xi=1$ for arbitrary atomic state. In the limit of $N\rightarrow\infty$, from Eq.~(\ref{limit2}), the Hamiltonian $H_{\rm I}$ reduces to the effective Hamiltonian
\begin{equation}
H_{\rm eff}=\lambda (b^\dagger a +ba^\dagger),
\end{equation}
where $b$ and $b^\dagger$ describes the effective bosonic mode. The above Hamiltonian describes the interaction of two bosonic modes, and can be used to construct a SWAP gate for two modes as follows~\cite{WangSwap} 
\begin{equation}
U_{\rm SWAP}=e^{i\frac{\pi}{2}(a^\dagger a +b^\dagger b)}~e^{-i\frac{\pi}{2}(b^\dagger a+ba^\dagger)}.
\end{equation}
Thus, the unitary operator $\exp(-iH_{\rm eff} t)$ at scaled time $\tau=\tau_0=\pi/2$ ($\tau=\lambda t$) is the SWAP gate up to the local operator $\exp[i{\pi}/2(a^\dagger a +b^\dagger b)]$.
We see that in the limit of $N\rightarrow\infty$, at time $\tau=\tau_0$, the state in one system swaps with that of anther system. Therefore, squeezing is of course perfectly transferred. 

\begin{figure}
\includegraphics[width=0.45\textwidth]{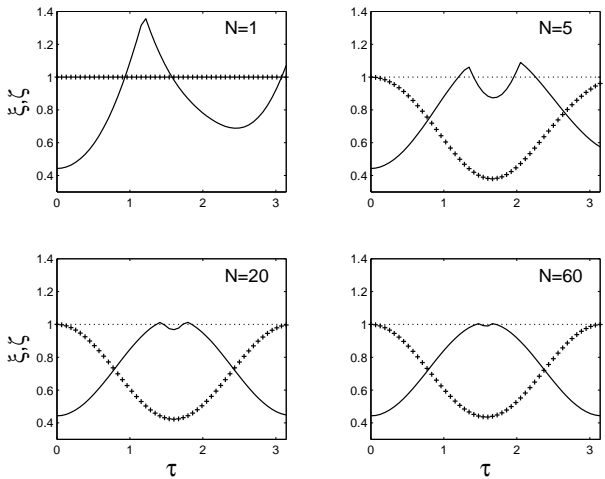}
\caption{The squeezing parameters $\zeta$ (solid line) and $\xi$ (cross points) versus $\tau$ for different number of atoms.}
\end{figure}

Now we numerically examine the intermediate case, i.e., $1<N<\infty$. Figure 1 plots the two squeezing parameters $\zeta$ and $\xi$ for state $|\psi(t)\rangle$. We observe that beside the case of $N=1$ (this case is plotted for comparison), the atomic state is spin squeezed over most of the time, and is maximally squeezed near the point $\tau=\tau_0$. 
We see a dip of bosonic squeezing around $\tau=\tau_0$, and it disappears with the increase of number of atoms. The dip is due to a complicated entanglement between light and atomic systems.  
As time goes on, for the case of $N=60$, the decrease (increase) of light squeezing goes with the increase (decrease) of atomic squeezing for most of the time, i.e, the transfer of squeezing from light to atoms is nearly perfect. 

\begin{figure}
\includegraphics[width=0.45\textwidth]{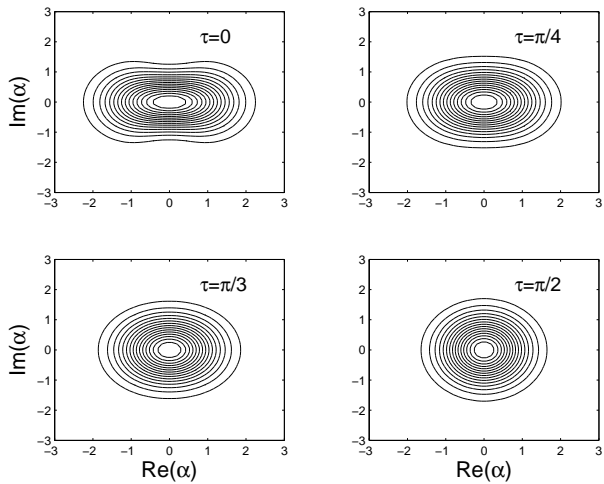}
\caption{Dynamics of the $Q$ function for the light field with $N=10$ and $\alpha=\alpha_0$.}
\end{figure}

To more clearly display the squeezing effect and the squeezing transfer, we plot the $Q$ function for the field in Fig.~2 and the Husimi ${Q}$ function for the atoms in Fig.~3. The two functions are given by
\begin{equation}
Q(\alpha)=|\langle\alpha|\psi(t)|^2, \quad {Q}(\eta)=|\langle\eta|\psi(t)|^2.
\end{equation}
From Fig.~2, we see clearly that the initial light squeezing due to quantum interference between two components~$|\pm\alpha\rangle$~in the EBCS. As times goes on, within the considered time range, light squeezing decreases. Contrary to the behaviour of the $Q$ function for field, from the dynamics of the Husimi function in Fig.~3, the atoms become more squeezed as time goes on. The $Q$ function at time $\tau=0$ and the Husimi function at $\tau=\pi/2$ are very similar, indicating near perfect transference of squeezing.

\begin{figure}
\includegraphics[width=0.45\textwidth]{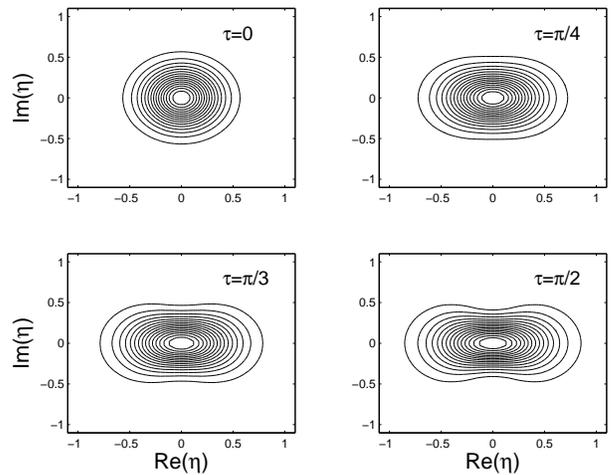}
\caption{Dynamics of the Husimi $Q$ function for atoms with $N=10$ and $\alpha=\alpha_0$.}
\end{figure}

\section{Discussion and conclusion}
In conclusion, we have examined the bosonic quadrature squeezing and atomic spin squeezing defined by Kitagawa and Ueda~\cite{Kitagawa}, and found that atomic squeezing reduces to bosonic squeezing for even and odd states. 
We have proved that the ESCS always exhibits spin squeezing, whereas the OSCS does not, which is similar to the case of bosonic squeezing that EBCS always exhibit quadrature squeezing, while the OBCS does not. For arbitrary atomic states, Eq.~(\ref{squ2}) is applicable since we can always rotate the mean spin  along $z$ direction. Then, formally the reduction of atomic squeezing to bosonic squeezing in the large $N$ limit can be obtained for arbitrary states. However, the rotation changes mean excitation number $\langle{\cal N}\rangle$. Thus, only when $\langle{\cal N}\rangle$ is low enough after the rotation can we take the limit and obtain the reduction.

Obviously, the reduction depends on the choice of criteria for squeezing. Now, we examine another spin squeezing criterion given in Refs.~\cite{Wineland,Sorensen01}, and see if it can reduces to the bosonic quadrature squeezing. The corresponding squeezing 
parameter for even and odd states can be written as~\cite{Wineland}
\begin{equation}
\xi^\prime =\frac{2j(\Delta S_{\mathbf{n}_{\bot }})^2}{|\langle S_z\rangle|^2}
=\frac{\xi}{|\langle S_z/j\rangle|^2}.
\label{xixixixi}
\end{equation}
We see that these two squeezing parameters, $\xi$ and $\xi^\prime$, differ only by a multiplicative factor. In the limit of $j\rightarrow\infty$, from Eq.~(\ref{limit2}), the squeezing parameter $\xi$ reduces to $\zeta$ (\ref{squu1}) for bosonic squeezing and the multiplicative factor reduces to 1. Then, from Eq.~(\ref{xixixixi}), the squeezing parameter $\xi^\prime$ also reduces to the bosonic squeezing parameter. Thus, according to the two typical spin squeezing definitions, the spin squeezing reduces to the bosonic squeezing in the limit of $j\rightarrow\infty$.

We have further studied squeezing transfer from light field to atomic system via the resonant Dicke Hamiltonian, and observe a nearly perfect transfer of squeezing at a certain interaction time for large number of atoms. In the context of experiments on spin squeezing of atomic ensembles, in general, there is large number of atoms. 
Therefore, we expect experimental observations of perfect squeezing transfer and light quadrature squeezing behaviours of atomic spin squeezing when the number of  atoms are large and the number of excited atoms are much less than the number of total atoms.

\acknowledgments
This project has been supported by an Australian Research Council Large Grant. 

\appendix
\section{calculations of squeezing for general states}

Although the squeezing parameters $\zeta $ of (\ref{squu1}) and $\xi $ of (\ref{squ2}) are derived for even and odd states, they are applicable to arbitrary states. Consider a state of light $|\Psi \rangle $ and the displaced state $|\Psi \rangle _D=D(\alpha )|\Psi \rangle,$ where $D(\alpha)=\exp(\alpha a^\dagger-\alpha^*a)$ is the displacement operator. The squeezing of these two states is identical as we have
\begin{equation}
D^{\dagger }(\alpha )X_\theta D(\alpha )=X_\theta +\alpha e^{-i\theta
}+\alpha ^{*}e^{i\theta },
\end{equation}
The operator $D^{\dagger }(\alpha )X_\theta D(\alpha )$ and $X_\theta $ only
differ by an additive constant; thus their variances on a state are equal and 
states $|\Psi \rangle _D$ and $|\Psi \rangle $ exhibit same squeezing.
Therefore, we can always displace a state $|\Psi \rangle $ by a quantity $-%
\tilde{\alpha}$ to adjust the expectation value of $X_\theta $ to zero and
keep the squeezing invariant, where $\tilde{\alpha}=\langle \Psi| a|\Psi
\rangle .$ Applying Eq.~(\ref{squu1}) to state $D(-\tilde{\alpha})|\Psi\rangle$ leads to 
\begin{equation}
\zeta=1+2(\langle a^\dagger a\rangle -|\langle a\rangle|^2)-2|\langle a^2\rangle-\langle a\rangle^2|, \label{psqu}
\end{equation}  
which is just the result in Ref.~\cite{Psqu}. 

For the atomic system, we consider the state of the atomic ensemble $|\Phi \rangle $
and rotated state 
\begin{equation}
|\Phi \rangle_R=R(\theta ,\phi )|\Phi \rangle, \; R(\theta,\phi)=e^{-\frac{\theta}2 \left(S_+e^{-i\phi}-S_-e^{i\phi}\right)}.
\end{equation}
The spin squeezing is invariant under the rotation $R(\theta ,\phi
)$~\cite{Usha03}, which we show briefly here.
Note the fact that the mean spin  $\langle\mathbf{S}\rangle$ for the two states  $|\Phi \rangle $ and $|\Phi \rangle_R$ differ by an orthogonal
matrix ${\cal R}$ 
\begin{equation}
_R\langle \Phi |\mathbf{S}|\Phi \rangle _R=\mathcal{R}\langle \Phi |\mathbf{S%
}|\Phi \rangle .
\end{equation}
Thus, the mean spin direction $\mathbf{n}$ and $\mathbf{n}_{\bot }$ for
state $|\Phi \rangle $ will become $\mathcal{R}\mathbf{n}$ and $\mathcal{R}
\mathbf{n}_{\bot }$ for the rotated state $|\Phi \rangle _R.$ 
The
absolute value of the mean spin is unchanged. 
From Eq.~(\ref{xixixi}), we may rewrite the squeezing parameter as 
\begin{equation}
\xi =\frac{2\langle \mathbf{Sn}_{\bot }^T\mathbf{n}_{\bot }
\mathbf{S}^T\rangle}j.
\end{equation}
Then, it becomes obvious that the squeezing parameters are the same for the two states $|\Phi \rangle $ and $|\Phi\rangle _R.$ As the spin squeezing is invariant under the rotation $
R(\theta ,\phi ),$ we can always rotate the mean spin along the $z$
direction by applying $R(-\tilde{\theta},\tilde{\phi})$ to state $|\Phi
\rangle $, and Eq.~(\ref{squ2}) becomes applicable for arbitrary spin states, where $\tilde{\theta}$ and $\tilde{\phi}$ can be determined from the mean spin of the state $|\Phi \rangle $ by $\tilde{\theta}=\arccos (\langle S_z\rangle
/|\langle \mathbf{S}\rangle |)$ and $\tilde{\phi}=\arctan (\langle
S_x\rangle /\langle S_y\rangle ).$

\end{document}